\documentstyle[12pt,bezier,psfig,pictex]{article}
\newcommand{\be}{\begin{equation}}
\newcommand{\ee}{\end{equation}}

\newcommand{\rf}[1]{(\ref{#1})}
\newcommand{\beq}{\begin{equation}}
\newcommand{\eeq}{\end{equation}}
\newcommand{\bea}{\begin{eqnarray}}
\newcommand{\eea}{\end{eqnarray}}

\renewcommand{\b}{\beta}
\renewcommand{\a}{\alpha}
\newcommand{\n}{\nu}
\newcommand{\m}{\mu}

\newcommand{\th}{\theta}

\newcommand{\om}{\omega}

\newcommand{\ra}{\right\rangle}
\newcommand{\la}{\left\langle}

\newcommand{\noi}{\vspace{12pt}\noindent}
\newcommand{\nn}{\nonumber \\}

\begin{document}
\topmargin 0pt
\oddsidemargin 5mm
\headheight 0pt
\headsep 0pt
\topskip 9mm

\hfill    NBI-HE-96-22

\hfill June 1996

\begin{center}
\vspace{24pt}
{\large \bf Alternative actions for quantum gravity \\
and the intrinsic  rigidity of the spacetime}

\vspace{24pt}

{\sl J. Ambj\o rn } and {\sl G.K. Savvidy}\footnote{Permanent address:
Yerevan Physics Institute, 375036 Yerevan, Armenia
}

\vspace{6pt}

 The Niels Bohr Institute\\
Blegdamsvej 17, DK-2100 Copenhagen \O , Denmark\\

\vspace{12pt}

{\sl  K.G. Savvidy}\\
 Princeton University, Department of Physics\\
 P.O.Box 708, Princeton, New Jersey 08544, USA

\end{center}
\vspace{24pt}

\vfill

\begin{center}
{\bf Abstract}
\end{center}

\noi
Using the Steiner-Weyl expansion formula for parallel manifolds and
the so called gonihedric principle we find  a large
class of discrete integral invariants which are defined on
simplicial manifolds of various dimensions. These integral
invariants include the discrete version of the Hilbert-Einstein
action found by Regge and  alternative actions which are  linear
with respect to the size of the manifold. In addition the concept of
generalized  deficit angles appear in a natural way
and is related to  higher order curvature terms.
These angles may be used
to introduce various aspects of rigidity in simplicial quantum gravity.

\vfill

\newpage

 \section{Introduction}

String theory and theories beyond string theory have increased the
interest for quantum gravity and physics at the Planck scale.
One problem in these theories is the lack of a non-perturbative
definition. Although string duality seems to open tantalizing
possibilities for extracting non-perturbative physics from
different perturbative sectors of the theory, it is still not
clear that we will be able to address the full physical  content
at the Planck scale without a non-perturbative definition of
string theory or its generalization. In particular, it is not clear
how much we will be able to say about the aspects related
to the quantum gravity sector of the theory.

It is therefore possible that quantum gravity, or at least important aspects
of quantum gravity, may be described entirely within the framework
of a non-perturbative field theory. A first step in this
direction is a non-perturbative definition
of the field theory we will denote quantum gravity. This is a non-trivial
task because the  continuum theory  has to be
invariant under reparametrizations.
Lattice gauge theory is an example of a successful
non-perturbative regularization of a quantum field theory
with a continuous internal symmetry. The regularization
breaks Euclidean invariance
(which is restored in the scaling limit), but maintains from a certain
point of view the concept of local internal invariance. In the case of gravity
the situation is  more difficult since we deal with local symmetries
of space--time itself, and any lattice regularization will break this
symmetry. In classical gravity a very natural discretization was
suggested by Regge \cite{regge} by the restriction to
piecewise linear manifolds,
and he showed how the Einstein--Hilbert action had a natural geometric
representation on the class of piecewise linear manifolds and could
be expressed entirely in terms of intrinsic invariants of the
piecewise linear manifolds. In this way one achieved a coordinate free
description of this class of manifolds and their actions, where the
dynamical variables were the link length.
The use of Regge calculus as a prescription for quantum gravity
is less straightforward, since there is not a one-to-one
correspondence between the piecewise linear metric and the length
assigned to the links (for a discussion, see
\cite{ninomiya} and also \cite{hamber} and references therein).
However, it is very encouraging that a variant of Regge calculus,
known as dynamical triangulation \cite{adf,david,kkm},
works perfect in two-dimensional quantum gravity.
In this formalism one fixes the link--length of the piecewise linear
manifold, and the assignment of a metric depends only on the connectivity of
the triangulation.
The summation over triangulations with different connectivity takes
the role of integration over inequivalent metrics and the action assigned to
the manifold is  calculated by Regge's prescription for a piecewise
linear manifold. In this formalism the scaling limit agrees with
the known continuum results of two-dimensional quantum gravity
\cite{kpz,ddk}, i.e. one gets
a reparametrization invariant theory in the scaling limit.
While it is easy to generalize the definition of
the two-dimensional discrete model
to both three \cite{av} and four dimensions \cite{aj}
(see also \cite{weingarten} for an earlier slightly different formulation),
the models can presently only be analyzed by numerical methods
\cite{many} and we have no continuum theory
of Euclidean gravity in dimensions larger than two
with which we can compare. Whether one
uses the formalism of dynamical triangulations or the original
formalism of Regge with fixed connectivity and variable
link length, it might well
be that the simplest versions of discretized Einstein--Hilbert action
which have been used so far at the discretized level are insufficient
in producing an interesting continuum limit in dimensions larger
than two. On the other hand it is very appealing to use
some class of piecewise linear manifolds as the natural choice
of discretization in quantum gravity since there is a one-to-one
correspondence between piecewise linear  structures and
smooth  structures for manifolds of dimensions less than seven.
This motivates the search for "natural" integral invariants which are
defined on piecewise linear manifolds.

In this article we would like to advocate a geometrical way to
construct integral invariants on a simplicial manifold.
It contains Regge's result as a special case and
allows us to construct a large
class of new integral invariants which might be helpful
in the attempts to define a regularized
path integral in quantum gravity which possesses an interesting
continuum limit.

The method is based on Steiner's idea of a {\it parallel manifold}
\cite{steiner,weyl,ambarzum}. Let $M_{n-1}$ be a smooth ($n$--1)-dimensional
manifold embedded in $n$-dimensional flat space $E^n$,
and let $M^\rho_{n-1}$ denote the so called parallel manifold defined
by displacing each point $P$ the distance $\rho$ along the outward normal
at $P$. The hyper-volume of the parallel manifold
$M^{\rho}_{n-1}$ can be expanded into a polynomial of the distance
$\rho$ from the original manifold . The coefficient to $\rho^k$
in this expansion
represents an integral invariant $\mu_{k}(M_{n-1})$, $k=0,1,...,n$--1
constructed on the ($n$--1)-dimensional manifold $M_{n-1}$ by means of the
the intrinsic and extrinsic curvature  \cite{weyl}.
It turns out that this expansion makes perfect sense
not only for smooth manifolds but also for a piecewise linear manifold
and the classical invariants $\mu_k(M_{n-1})$ become natural invariants
also for simplicial manifolds, in the same way as the integral
of the scalar curvature in the classical work of Regge has
a natural geometric definition directly on the piecewise linear manifolds.
In fact the Einstein--Hilbert action is equal $\m_2(M_{n-1})$ and
therefore among the classical invariants.

An important observation emerging from the  Steiner expansion
for piecewise linear manifolds is that
in all cases the integral invariants $\mu_{k}(M_{n-1})$
are the product of the $volume$ of the faces of $M_{n-1}$
and the $volume$ of the corresponding {\it normal images } or
{\it spherical image}  of these faces
(to be defined precisely in the next section):
\be
{\rm integral~invariant}= \sum_{\{faces\}} ({\rm volume~of~face} ) \cdot
({\rm volume~of~spherical~image}) \label{fact}.
\ee
In this equation the notation "face" means vertex, edge, triangle, tetrahedron
and higher dimensional sub-simplexes of $M_{n-1}$.

The generalization of the factorization property (\ref{fact}) of the
classical integral invariants in the form of gonihedric principle
allows to construct new integral invariants on a manifold
$M_{n-1}$. In accordance with this principle one should always
multiply the {\it volume} of the face by one of the
{\it geometric  measures}, i.e. length, area, volumes etc, associated with
the corresponding spherical image:
\be
\sum_{\{faces\}} ({\rm volume~of~face} ) \cdot
({\rm geometric~measure~on~spherical~image}). \label{goni}
\ee
This form of extension of classical integral invariants (\ref{fact})
maintain the locality of the integral invariants and
allow  to construct a large class of discrete integral
invariants defined on triangulated manifolds.

Performing a duality transformation it is possible to obtain a dual form of
the new invariants
\be
\sum_{\{faces\}} ({\rm geometric~measure~on~face} ) \cdot
({\rm volume~of~spherical~image}). \label{goni1}
\ee
It is also useful to consider more general integral invariants,
obtained from \rf{goni1} by replacing the volume density of the
spherical image with a function $\theta({\rm volume~on~image})$
of the volume density of the spherical image,
where the choice of function $\theta$ will be dictated
by the specific physical problem under consideration \cite{savvidy}.

Since the spherical images are defined by  the embedding of
the manifold into Euclidean  space it is not an easy task
to understand why some of the integral invariants are nevertheless
intrinsic and independent of the embedding. In fact, Gauss himself
was very surprised when he discovered that the ``Gaussian curvature"
which, in analogy with the curvature of a path, was defined by
means of the spherical image, was indeed a {\it bending invariant}.
This is a ``Theorema egregium'', a ``most excellent theorem'', wrote
Gauss. In modern language we understand that the underlying reason
is that for smooth manifolds some of the
integral invariants are constructed from the Riemann tensor and thus
are independent of their embedding in Euclidean space. For the
discrete version of the classical invariants and for the new ones
considered here this can be seen applying the  Gauss-Bonnet theorem to the
spherical images \cite{herglotz,allendoerfer,santalo,schneider}
and expressing the measures on the image in terms of  the internal angles .

In particular, in four dimensions, this approach allows to obtain
natural discrete
representation of the gravitational action which contains terms
quadratic in curvature tensor. The Hilbert-Einstein action
together with the higher derivative terms take the form
\bea
Action &=& \frac{1}{G}
\sum_{\la ijk\ra } \sigma_{ijk} \cdot \omega^{(2)}_{ijk}+g_1 \sum_{\la ijk\ra}
\om^{(2)}_{ijk} +g_2\sum_{\la ijk\ra}\Bigl(\om^{(2)}_{ijk}\Bigr)^2 + 
\label{higherdev}\\
&&
+ k_1 \sum_{\la i\ra } ( 1 - \frac{1}{2\pi^{2}}  \Omega^{(4)}_{i})+
k_2  \sum_{\la i\ra }
\Bigl( 1 - \frac{1}{2\pi^{2}} \Omega^{(4)}_{i}\Bigr)^2
 +\cdots,
\nonumber
\eea
where $\sigma_{ijk}$
is the area of the triangle $<ijk>$,  $\omega^{(2)}_{ijk}$ is the
deficit angle associated with the same triangle and
the first term is just the Regge action of the piecewise linear
manifold.  The terms associated with the coupling constants $g_i$
are functions of the deficit angles  $\omega^{(2)}_{ijk}$ and
in this way they corresponds to higher derivative terms.
The quantity $\Omega^{(4)}_i$ is the total solid angle 
associated with the vertex $i$, i.e.
$\Omega^{(4)}_i = \sum \Omega^{(4)}_{ijklm}$, where $\Omega_{ijklm}^{(4)}$
  is the solid angle at the vertex of the
four dimensional tetrahedron $\la ijklm\ra$ and the   summation is over 
adjacent tetrahedra. The terms associated with the $k_i$'s
represent  functions of the  ``solid deficit angles'' at each vertex $i$.
As we shall see these terms are likewise related to higher derivative terms.
In a discretized approach to quantum gravity, where the
functional integration is first defined  by integration
over a suitable set of piecewise linear manifolds, and the continuum limit
is taken afterwards, it is
natural to include these terms. They will introduce
an intrinsic  rigidity into simplicial quantum gravity,
which might be needed if one should be able to define an interesting
continuum limit simplicial quantum gravity.

In the next section we shall review the Steiner-Weyl expansion
method and derive the discrete version of the classical invariants.
In subsequent sections we shall construct new extensions of these
functionals in two, three, four and high dimensions and apply this
approach to get discrete version of the high derivative terms in
quantum gravity.

\section{Integral invariants on smooth manifold}

Let $M_{n-1}$ be a compact orientable hyper-surface
embedded in an Euclidean space $E^{n}$ by a map $P \mapsto X_\m(P)$.
At a point $P$
of $M_{n-1}$ there are $n$--1 principal curvatures $R_{i}$,
$i=1,.., n$--1. If $d v_{n-1}$ denotes the element of hyper-volume
of $M_{n-1}$, the integral invariants of
$M_{n-1}$ can be defined as a integrals over symmetric functions of
the principal radii of curvature $R_{i}$ \cite{weyl}
\be
\mu_{k}(M_{n-1}) \equiv \int \{\frac{1}{R_{i_{1}}}...\frac{1}{R_{i_{k}}} \}
d v_{n-1} \label{inva},
\ee
where $k=0,1,..,n$--1 and $\{...\}$ denotes the symmetrization .
In particular, $\mu_{0}$ is the hyper-volume, $\mu_{2}$ is the
Hilbert-Einstein action
and $\mu_{n-1}$ is the  {\it degree} of the so called normal map
$$
P \in M_{n-1} \mapsto n(P)\in S^{n-1},
$$
which maps a point $P$ of the manifold $M_{n-1}$
into the unit vector $n(P)$ normal to $M_{n-1}$. Let $\Omega$
be a subset of $M_{n-1}$. Then $n(\Omega)$ is called
the {\it spherical image} of $\Omega$.
If $d \om_{n-1}$ denotes the  hyper-volume element
on $S^{n-1}$, then
\be
d v_{n-1} = R_{1}...R_{n-1} d \omega_{n-1} \label{volua}
\ee
where the product $1/(R_{1}...R_{n-1})$ is the Gauss-Kronecker
curvature. This is the generalization of the famous two-dimensional
formula and one can use this formula to express the integral invariants
\rf{inva} of  $M_{n-1}$ as integrals over the spherical image of $M_{n-1}$:
\be
\mu_{k}(M_{n-1}) = \int \{ R_{i_{1}}...R_{i_{n-k-1}} \}
d \omega_{n-1} \label{invar},
\ee
where $k=0,1,...,n$--1. It is now easy to see  that
$\mu_{n-1}$ is the degree of map $n: M_{n-1} \mapsto S^{n-1}$
defined by the field of normals.

The integral invariants (\ref{inva}) and  (\ref{invar}) appear in a
natural way in the so called {\it Steiner-Weyl expansion formula} for the
hyper-volume of the parallel manifold $M^{\rho}_{n-1}$
\cite{steiner,weyl}. If $M_{n-1}$ is embedded in $E^{n}$ we define
a parallel manifold $M^{\rho}_{n-1}$ as a set of all
points at a distance $\rho$ from $M_{n-1}$, i.e. by the map
$P \mapsto X_\m(P)+\rho \cdot n_\m(P)$,
where $n(P)$ again denotes the outward normal at $P$.
For $\rho$ sufficiently small the hyper-volume $\mu_{0}(M^{\rho}_{n-1})$
of the parallel manifold is equal to
\be
\mu_{0}(M^{\rho}_{n-1}) = \int  (R_{1}+\rho)...
(R_{n-1}+\rho)
d \omega_{n-1} \label{expa},
\ee
simply because the $n$--1 principal curvatures for the parallel
manifold $M^{\rho}_{n-1}$ are equal to $R_{i}+\rho$,~~~$i=
1,2,..,n$--1. Expanding the product of the integrand we get
\be
\mu_{0}(M^{\rho}_{n-1}) =\mu_{0}(M_{n-1}) +\rho\cdot \mu_{1}(M_{n-1})
+\rho^{2}\cdot \mu_{2}(M_{n-1})+...+ \rho^{n-1}\cdot \mu_{n-1}(M_{n-1})
         \label{invari},
\ee
thus generating the whole sequence of integral invariants which we
discussed above.

It can be shown that the  coefficients $\m_{2k} (M_{n-1})$ to the
even powers of $\rho$ in the expansion \rf{invari} are independent
of the embedding \cite{weyl}, while the coefficients $\m_{2k+1} (M_{n-1})$
to the odd powers of $\rho$ refer explicitly to the extrinsic geometry.

\section{Integral invariants on piecewise linear manifolds}

The idea of a parallel manifold allows us to define the discrete
versions of the integral invariants (\ref{inva}) and  (\ref{invar})
for piecewise linear manifolds and in this way generalize the
work of Regge.

\begin{figure}
\centerline{\input{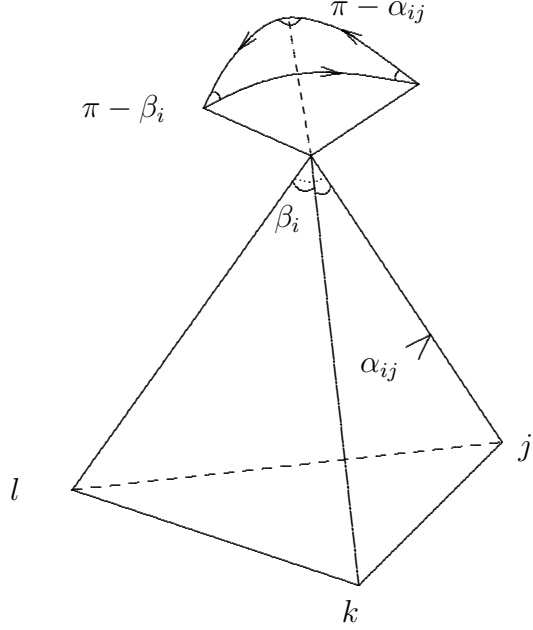}}
\caption[fig1]{The spherical image of the vertex $\la i\ra$. The
internal angle between edges are denoted as $\b_i$ and the
external angles between triangles as $\a_{ij}$.}
\label{fig1}
\end{figure}

First we consider two-dimensional surfaces \cite{ambarzum}.
Let $D_{3}$ denote a three-dimensional  piecewise linear
manifold in $E^3$ with a  connected boundary.
The boundary of $D_3$ will be a piecewise linear surface $M_2$ embedded in
$E^3$.  We define angles between
edges (one simplexes) $\la  ij \ra $ and $\la ik\ra $ as $\beta_{ij;ik}$
and between triangles (two simplexes) $\la ijl\ra $ and $\la ijk\ra $
as $\alpha_{ijl;ijk}$. These angles completely define the
internal and external geometry of a surface. It is easier
to use the shorthand  notation
\begin{eqnarray}
\beta_{ij;ik} \equiv \beta_{i} &~~~~~&
{\rm for~internal~angles} \label{ang}\\
\alpha_{ijl;ijk} \equiv \alpha_{ij} &~~~~~ &{\rm for~external~angles}
\label{angl}
\end{eqnarray}
and if needed one can recover the whole notation, see Fig.\ \ref{fig1}.

Note that the  parallel surface $M^{\rho}_{2}$ is
{\it one time differentiable} even if $M_{2}$ is only piecewise linear.
The area of $M^{\rho}_{2}$ contains separate parts which we shall
compute. The first part, $S(M_{2})$, is equal to the area of the
original surface $M_{2}$, i.e. it originate from parallel displacement of
the triangles $\la ikj \ra$ constituting $M_2$ the distances
$\rho \cdot n (\la ijk \ra)$, where $n (\la ijk\ra )$ denotes the
outward normal of triangle $n (\la ijk \ra)$ in the orientable
triangulation of $M_2$:
\be
S(M_2) = \sum_{\la ijk\ra } \sigma_{ijk}\cdot 1 \label{area},
\ee
where $\sigma_{ijk}$ is the area of the triangle $\la ijk\ra $
and the summation is extended over all triangles of the surface
$M_{2}$. The second part of the area, $\rho A(M_{2})$,
of the parallel surface
is equal the sum of the areas of the cylinders surrounding
the edges \cite{ambarzum}, i.e. proportional to the displacement $\rho$,
and we can write
\be
\rho \cdot A(M_2) = \rho \cdot \sum_{\la ij\ra } \lambda_{ij}
\cdot (\pi-\alpha_{ij}) \label{line} ,
\ee
where $\lambda_{ij}$ is the length of the edge $\la ij\ra $ and
$\alpha_{ij}$ is the dihedral angle at the edge $\la ij\ra $ and the
summation is over all edges. The third part of the area
of $M^{\rho}_{2}$, $\rho^{2} \chi(M_{2})$,
is equal to the sum of the areas of the
spherical polygons which surround the vertices of $M_{2}$, i.e. proportional
to $\rho^2$, and we can write
\be
\rho^2 \cdot \chi(M_2) =\rho^{2}\cdot \sum_{\la i\ra } 1\cdot
(2\pi -\sum\beta_{i}) \label{topo},
\ee
where $\chi_{i} = 2\pi -\sum\beta_{i}$ is the area of the
spherical polygon on a unit sphere corresponding to a vertex
$\la i\ra $, usually called  a deficit angle,
and the summation is over all vertices of $M_{2}$.

Comparing these quantities with the  Steiner expansion
(\ref{invari}) for smooth two-dimensional surfaces
\be
S(M^{\rho}_{2}) = \int_{M_{2}}R_{1}R_{2} d\omega
+ \rho \cdot\int_{M_{2}}(R_{1} + R_{2}) d\omega +
\rho^{2} \cdot \int_{M_{2}} d\omega ,
\ee
we can get the natural discrete representation of the integral
invariants for the two-dimensional surface \cite{savvidy1}
\begin{eqnarray}
S(M_{2}) &=& \sum_{\la ijk\ra } \sigma_{ijk}\cdot 1,\label{area12}\\
A(M_{2}) &=& \sum_{\la ij\ra }\lambda_{ij}\cdot ( \pi - \alpha_{ij}),
\label{line13}\\
\chi(M_{2}) &=& \sum_{\la i\ra }1\cdot ( 2\pi - \sum \beta_{i}), \label{disc}
\end{eqnarray}
see Fig.\ \ref{fig1}.

\section{Factorization and the gonihedric principle}
The important message we get from the Steiner expansion
and from the formulae above is that
the integral invariants are (the sum of) products of a
volume of the faces of the surface $M_{2}$
and a volume of the spherical images of these faces
\be
{\rm integral~invariant}= \sum_{\{faces\}} ({\rm volume~of~face} ) \cdot
({\rm volume~of~spherical~image}) \label{stru}.
\ee
Indeed the spherical images of the faces of $M_{2}$ are:
\begin{eqnarray}
{\rm triangle}~~ \la ijk\ra  &\rightarrow&  {\rm point~on~}S^{2},
\label{trian16} \\
{\rm edge}~~~ \la ij\ra  &\rightarrow& {\rm arc~on~}S^{2},
\label{edge17}\\
{\rm vertex}~~~~ \la i\ra &\rightarrow& {\rm spherical~polygon~on~}S^{2}.
\label{imag}
\end{eqnarray}
and the volume elements on these spherical images are
\bea
{\rm triangle}~\la ijk\ra  &\rightarrow &1
\label{trian19}\\
{\rm edge}~~\la ij\ra & \rightarrow& (\pi-\alpha_{ij})
\label{edge20}\\
{\rm vertex}~ \la i\ra  &\rightarrow & (2\pi - \sum \beta_{i}),
\label{vert21}
\eea
i.e. equal to one for a point in $S^2$ which is the image of a triangle,
equal to the
length of the arc in $S^2$ which is the image of an edge, and equal to
the area of the polygon on $S^2$ which is the image of a vertex.
They are functions of the internal and external angles (\ref{ang}),
and (\ref{angl}).
The factorization or goni-hedric structure (\ref{stru}) of
the integral invariants (\ref{area12})-(\ref{disc})
is transparent now.

The generalization of the factorization property \rf{stru} of the
classical integral invariants in the form of gonihedric principle
can be used now to construct new integral invariants on a manifold
$M_{2}$. In accordance with this principle one should always
multiply the {\it volume} of the face by one of the
{\it geometric  measures}, associated with
the corresponding spherical image:
\be
\sum_{\{faces\}} ({\rm volume~of~face} ) \cdot
({\rm geometric~measure~on~spherical~image}).
\label{stru1}
\ee

In the two-dimensional case considered so far
the area functional $S(M_2)$ is uniquely defined since there is
no nontrivial measure
associated with the point on $S^2$ which is the spherical image of the
triangles of $M_2$.  The same is true for the linear
functional $A(M_{2})$, because only the length of the arc
can be associated with the
spherical image (\ref{edge20}) of the edge (\ref{edge17}).

\begin{figure}
\centerline{\input{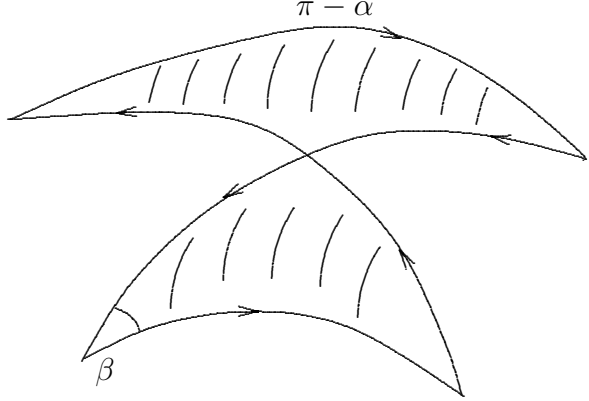}}
\caption[fig2]{On every spherical polygon one can define the area
\rf{area22}, the perimeter \rf{curv23} and the curvature \rf{mesu}.}
\label{fig2}
\end{figure}

Only the  topological invariant (\ref{disc}) can lead to new expressions
because the spherical image (\ref{imag})
of the vertex $\la i\ra $ is a polygon
on $S^{2}$ (see Fig.\ \ref{fig2}),
and the polygon has sufficiently
structure to allow non-trivial measures. The possible purely
geometric measures defined on a polygon on $S^2$ are (see Fig.\ \ref{fig2}):
\begin{eqnarray}
{\rm the~ area}~~~ &=& (2\pi-\sum \beta_{i}),\label{area22}\\
{\rm the~ curvature}& = &\sum (\pi -\beta_{i}),\label{curv23}\\
{\rm the~ perimeter} &=& \sum_{j} (\pi-\alpha_{ij}).
\label{mesu}
\end{eqnarray}

As we have seen the total area of the spherical image
is proportional to the Euler character
(see (\ref{disc}), (\ref{vert21}) and (\ref{area22}).
The total curvature of the spherical polygon {\it cannot be used},
because it is possible by a continuous deformation of $M_2$ to ensure
that there is zero curvature (no deficit angle) associated
with a given vertex, in which
case the polygon image (\ref{imag})
will disappear but it will  leave a  nonzero curvature (\ref{curv23}).
Such a situation is not physical interesting and we exclude it.
This means that only the perimeter (\ref{mesu})
can be used in the construction of new invariants,
and in accordance with (\ref{stru1}) we define
\be
\Theta(M_{2}) = \sum_{\la i\ra }\Bigl(\sum (\pi - \alpha_{ij})\Bigr).
\ee
Every term with dihedral angle $\alpha_{ij}$ appears in
both vertices $\la i\ra $ and $\la j\ra $
therefore after resummation one get
\be
\Theta(M_{2}) = 2\sum_{\la ij\ra }(\pi - \alpha_{ij}) \label{leng}.
\ee
The trivial resummation of quantities given on a vertices
into a sum over adjacent edges will again appear in  high
dimensions but in the less trivial form of a dual transformation.

The invariant (\ref{leng}) is equal to the total
length of the arcs of the spherical
image of all edges of the surface $M_{2}$. This
quantity is a natural definition of the total {\it extrinsic curvature}
of the triangulated surface.
There is a simple mnemonic rule by which one can get this
integral invariant out of the formulae already available,
i.e.\ (\ref{area12})-(\ref{disc}).
Indeed, if we shall consider a triangulated surface with fixed length
of all edges $\lambda_{ij}=a$ (as in the case of dynamical
triangulated surfaces), then
the functional $A(M_{2})$ is proportional to $\Theta(M_{2})$
\be
A(M_{2}) =a \cdot \Theta(M_{2}).
\ee
Let us finally consider generalizations  where
$$
{\rm measure~on~spherical~image } \to
\th({\rm measure~on~spherical~image})
$$
as discussed above. The obvious and simplest example is
$$
\pi -\a_{ij} \to |\pi -\a_{ij}|.
$$
In this way we can define a new integral invariant
\beq\label{new1}
\tilde{\Theta} (M_2) = \sum_{\la ij\ra }|\pi - \alpha_{ij}|,
\eeq
which is a natural discretized version of
\be
\int_{M_{2}}\left(\frac{1}{R_{1}}+\frac{1}{R_{2}}\right)^2 d S.
\ee
It is also possible to consider more general functions
\cite{savvidy}
\be
\pi -\alpha \rightarrow \theta(\alpha),~~~~~\theta(\pi)=0
\ee
where $\theta(\alpha)$ increases monotonically at the both sides
from the point $\alpha =\pi $ and near that point has the
following parametrization
\be\label{stru1a}
\theta(\alpha)=(\pi -\alpha)^{\varsigma}.
\ee
For these more general functions \rf{new1} is replaced by
\be\label{new2}
\tilde{\Theta} (M_2) = \sum_{\la ij\ra }\th (\alpha_{ij}).
\ee
The index $\varsigma$  (\ref{stru1a}) is important for the
convergence of the corresponding
partition function and the critical properties of the system
\cite{savvidy,durhuus}.

In order to demonstrate the value of these invariants for the theory of
random surfaces, let us compute them for a  sphere of radius $R$
which is triangulated in the following way: first we divide it
in $n$ degrees of latitude and $m$ degrees of longitude and
each square obtained this way
is divided in two triangles. With this division we  get
for the action $A(M_2)$ defined by \rf{line13}:
\be
A(S^{2};\varsigma) = 4\pi R \left((\frac{n}{a})^{1-\varsigma}
+(\frac{m}{b})^{1-\varsigma}\right)
\ee
and for the action $\tilde{\Theta}(M_2)$ defined by \rf{new2}:
\be
\tilde{\Theta}(S^{2};\varsigma)= a n m^{1-\varsigma} + b m n^{1-\varsigma},
\ee
where $a$ and $b$ are constants.  We have
different behavior of the action in the limit $n,m \rightarrow \infty$.
If $\theta(\a) = 1-\cos \alpha $, i.e.\ $\varsigma = 2$, we get
\be
\tilde{\Theta}(S^{2};\varsigma=2)=\pi^{2} \frac{m}{n} +2\pi \frac{n}{m}.
\ee
Thus the index $\varsigma$ influence in an essential way the convergence
of the partition function, and should be chosen less than one in order
to have an interesting theory.

\section{Three-dimensional manifolds}

We consider now a four-dimensional  domain $D_{4}$  in $E^4$
bounded by the three-dimensional manifold $M_{3}$. The
three-volume of the parallel manifold $M^{\rho}_{3}$ has the
form (\ref{invari}), i.e.
\bea
V(M^\rho_{3})& =& \int_{M_3} R_{1}R_{2}R_{3} d\omega_3
+\rho \cdot \int_{M_3} \{ R_{1}R_{2} + R_{2}R_{3} +R_{3}R_{1} \} d\omega_3
\nn
&&
+\rho^{2} \cdot \int_{M_3} \{ R_{1}+R_{2}+R_{3}\} d\omega_3
+\rho^{3} \cdot \int_{M_3} d\omega_3 \label{inva3a}
\eea
or in terms of notation  used so far
\be
V(M^{\rho}_{3})=V(M_{3}) + \rho \cdot S(M_{3}) + \rho^{2}\cdot A(M_{3}) +
\rho^{3} \cdot N(M_{3})
\label{inva3}
\ee
which defines the volume, the area, the linear size
and the topology of the domain.

We now apply the method of parallel expansion
to the  piecewise linear  three-dimensional manifolds
to get discrete versions of the
corresponding integral
invariants. Following the procedure in two dimensions we
introduce angles between one, two and three simplexes
in order to define the
internal and external geometry of the simplicial manifold (see Fig.\ 3):
\begin{figure}
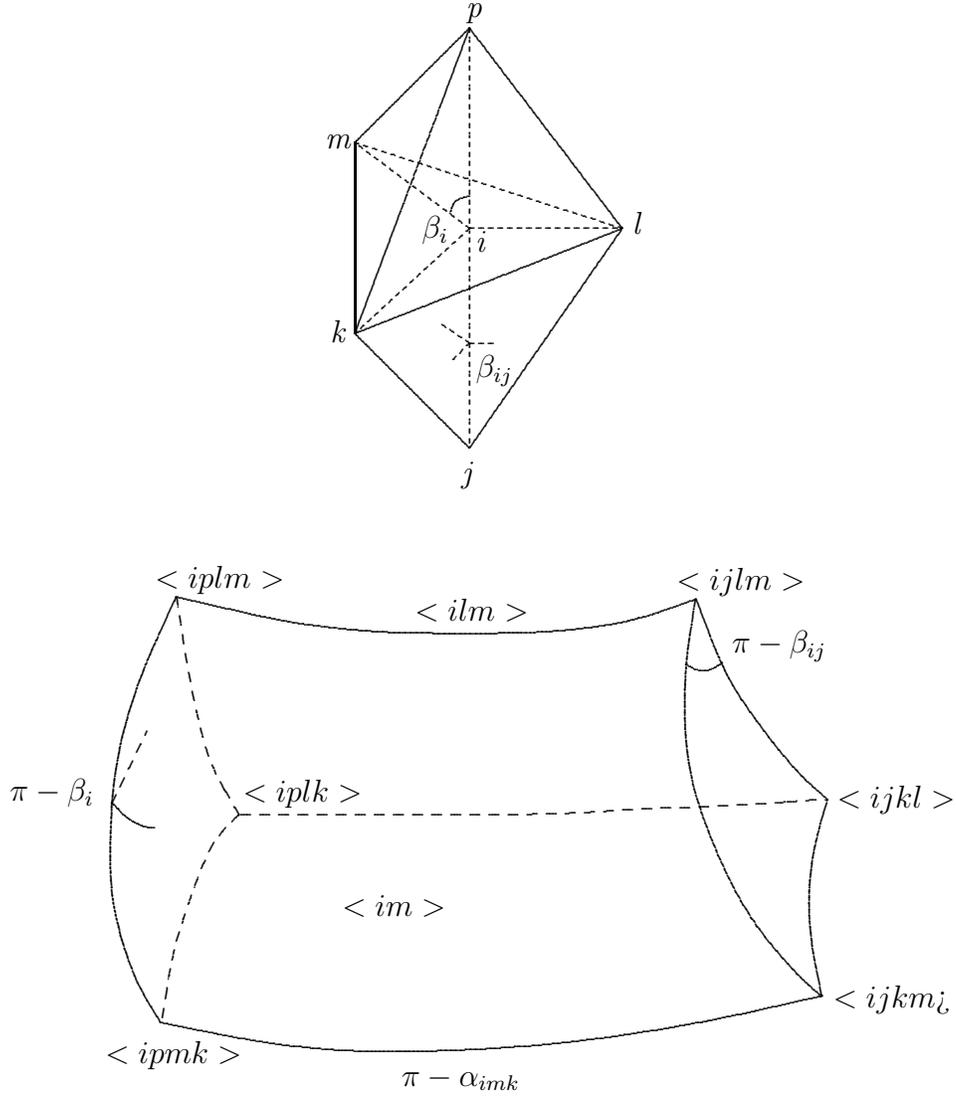

\vspace{-2cm}
\centerline{\input{Fig3a}}
\vspace{1cm}
\centerline{\input{Fig3b}}
\vspace{0.5cm}
\caption[fig3]{The vertex of the three-dimensional simplex (shown on the
upper figure) and the corresponding spherical polyhedron on $S^3$.
The volume of this image is denotes $\om^{(3)}_i$, the area as
$\sum \om^{(2)}_{ij}$, where the summation is over all polygons of the
given polyhedron, and finally the linear size as $\sum\Omega^{(2)}_{ijk}
\om^{(1)}_{ijk}$ and the summation is over all arcs of the polyhedron}
\label{fig3}
\end{figure}
\begin{eqnarray}
\beta_{ij;ik} \equiv \beta_{i}&~~~~& {\rm for~internal~angles} \label{a0}\\
\beta_{ijl;ijk} \equiv \beta_{ij}&~~~~& {\rm for~internal~angles}
\label{a01}\\
\alpha_{ijkl;ijkm} \equiv \alpha_{ijk}&~~~~&{\rm for~external~angles}
\label{a}
\end{eqnarray}
and we introduce
the corresponding spherical images of the simplexes of $M_{3}$ as
\begin{eqnarray}
{\rm tetrahedron~~}\la ijkl\ra~~ &\rightarrow& {\rm~~ point~on~}S^{3}
\label{34}\\
{\rm triangle~~} \la ijk\ra ~~&\rightarrow& {\rm ~~arc~on~}S^{3},
\label{35} \\
{\rm edge~~} \la ij\ra~~ &\rightarrow&{\rm ~~polygon~on~}S^{3},
\label{36}\\
{\rm vertex~~} \la i\ra ~~ &\rightarrow&{\rm ~~ spherical~polyhedron~on~}S^{3}
\label{image}
\end{eqnarray}
Using the angles already defined on the simplexes (\ref{a0})-
(\ref{a}) one can find the geometric measures on the spherical images
(\ref{34})-(\ref{image}):
\begin{eqnarray}
{\rm tetrahedron~~}\la ijkl\ra ~~&\rightarrow& ~~1
\label{38}\\
{\rm triangle~~}\la ijk\ra ~~&\rightarrow& ~~(\pi-\alpha_{ijk})
\label{39}\\
{\rm edge~~}\la ij\ra ~~&\rightarrow&~~(2\pi - \sum \beta_{ij})
\label{40}\\
{\rm vertex~~}\la i\ra ~~&\rightarrow&~~\omega^{(3)}_{i}(\alpha ,\beta)
\label{mesu3}
\end{eqnarray}
Using this information it is not difficult to compute the volume
of the parallel manifold. It will contain four parts which we
should identify with invariants (\ref{inva3}) of the simplex $M_{3}$.
If $v_{ijkl}$ denotes the volume of the tetrahedron $\la ijkl\ra$ and
$\om^{(3)}_i$ denotes the volume in $S^3$ of the spherical polyhedron
corresponding to vertex $\la i\ra$, we obtain:
\bea
{\rm the~volume~~}V(M_{3})~& = &~\sum_{\la ijkl\ra } v_{ijkl}\cdot 1
\label{xx1}\\
{\rm the~ area~~} S(M_3)~& = &~
\sum_{\la ijk\ra }\sigma_{ijk} \cdot (\pi - \alpha_{ijk} ),
\label{xx2}\\
{\rm the~ linear~ function~~}A(M_{3})~&=&~ \sum_{\la ij\ra }\lambda_{ij} \cdot
(2\pi - \sum \beta_{ij}),
\label{xx2a}\\
{\rm the ~topological~ invariant~~}
N(M_{3})~ &=& ~\sum_{\la i\ra } 1 \cdot \omega^{(3)}_{i}(\alpha,\beta).
\label{topo45}
\eea
All these integral invariants have the factorized form (\ref{stru}).

Let us now consider the possible new integral invariants along the lines
already discussed in the two-dimensional case.
Again the spherical images of the
two highest dimensional simplexes, the tetrahedron and the triangle
are of lowest dimension, i.e. a point and an arc, respectively,
and it is impossible to find any geometric extension for these
objects. The first non-trivial simplex is from this point of view
the edge, which has as its spherical image a polygon. For the polygon we can
in addition to its area also use its perimeter as a geometric measure,
and hence we find a linear invariant different from $A(M_3)$:
\be
\Lambda(M_{3})=\sum_{\la ij\ra }\lambda_{ij} \cdot
\Bigl(\sum (\pi - \alpha_{ijk})\Bigr),
\label{lamb}
\ee
in accordance with our principle (\ref{stru1}).
The spherical image of a vertex $\la i \ra$ is a spherical polyhedron
(\ref{image}), and for this object we have
the topological invariant (\ref{topo45}), which corresponds to
the use of the three-volume $\omega^{(3)}_{i}$
as the geometric measure. However, we can also use
the arc--length or the area of the two--simplexes in the polyhedron as
geometric measure (see Fig.\ \ref{fig3}), i.e.:
\be
\sum ~\beta_{i}\;(\pi - \alpha_{ijk})~~~~{\rm or}~~~~
\sum~(2\pi - \sum \beta_{ij}).
\ee
This leads to the following new integral invariants
\be
\Theta(M_{3}) = \sum_{\la i\ra } 1 \cdot \Bigl(\sum \beta_{i}\;(\pi -
\alpha_{ijk})\Bigr) \label{chi49}
\ee
\be
\Xi(M_{3}) = \sum_{\la i\ra } 1 \cdot \Bigl(\sum (2\pi -
\sum \beta_{ij})\Bigr).  \label{xi}
\ee

\subsection{The dual representation}
There is an equivalent or ``dual'' form of the invariants \rf{lamb},
\rf{chi49} and \rf{xi}, which can be obtained after resummation.
In (\ref{lamb}) we can combine all edges belonging to a given
triangle $\la ijk\ra $ to get a sum
$\lambda_{ij}+\lambda_{jk}+\lambda_{ki}=\lambda_{ijk}$.
Hence,
\be
\Lambda(M_{3}) = \sum_{\la ijk\ra }\lambda_{ijk} \cdot (\pi -\alpha_{ijk})
\label{lamb1}
\ee
where $\alpha_{ijk}$ is the angle between two neighboring tetrahedra
of $M_{3}$ having a common triangle $\la ijk\ra $ of the perimeter
$\lambda_{ijk}$. This is the
dual form of the integral invariant (\ref{lamb}) since  either one
can multiply the length of the edge by the total perimeter
of the polygon which is the spherical image of the edge, or
one can multiply the perimeter of the triangle by the length of the arc
of the spherical image of the triangle, as in (\ref{lamb1}).

The same kind of transformations, in the form of a resummation
of the quantities given at the vertices and in this way transferring the
sum to one over adjacent edges or triangles,
work for the two invariants (\ref{chi49}) and  (\ref{xi}).
Collecting the terms belonging to the
same triangle $\la ijk\ra $ in (\ref{chi49}) we can get
$(\beta_{i}+\beta_{j}+\beta_{k})
(\pi-\alpha_{ijk})= \pi (\pi-\alpha_{ijk})$ therefore
\be\label{extrinsic}
\Theta(M_{3})= \sum_{\la ijk\ra } (\pi-\alpha_{ijk}).
\ee
Note that $\Theta (M_3)$ is the total extrinsic curvature of $M_3$.
If we collect the terms belonging to the same edge in (\ref{xi})
we simply get
\be\label{deficit}
\Xi(M_{3})= \sum_{\la ij\ra } (2\pi-\sum \beta_{ij}).
\ee
$\Xi(M_3)$ is the total deficit angle of $M_3$.

The gonihedric principle (\ref{stru1})
which we used to construct extensions can now
be formulated in  $dual$ form: either one should
multiply the volume of the face by all possible geometric measures
which can be constructed on its spherical image,
or one should multiply all possible geometric measures on the face
by the volume of its spherical image.

Finally, we remark that also in the three-dimensional case
one can of course introduce more general
invariants by considering functions $\th$ of the geometric measures on the
spherical images, as was done in the two-dimensional case.

\section{Four-dimensional manifolds}

\begin{figure}
\centerline{\hbox{\psfig{figure=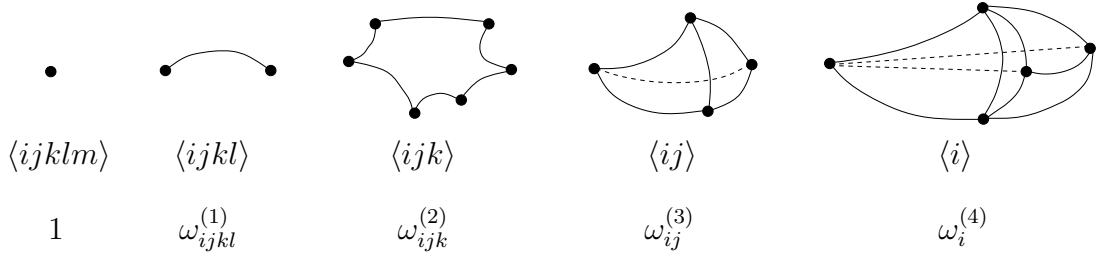,width=14cm}}}

\noindent
$\la ijklm\ra$ \hspace{0.7cm}$\la ijkl\ra$\hspace{1.7cm}
$\la ijk\ra$ \hspace{2.4cm}$\la ij\ra$\hspace{3.2cm}$\la i\ra$

\vspace{12pt}
\noindent~~~~1 \hspace{1.4cm}$\om^{(1)}_{ijkl}$\hspace{2.0cm}
$\om^{(2)}_{ijk}$\hspace{2.6cm}$\om^{(3)}_{ij}$\hspace{3.1cm}
$\om^{(4)}_i$
\vspace{24pt}
\caption[fig4a]{The sub-simplexes of the four-dimensional manifold
(second line)
and the corresponding spherical images on $S^4$ (first line).
The hyper-volume of the
image is defined as $\om^4_i$, the volume of the image as
$\sum \om^{(3)}_{ij}$,
the area of the image as $\sum \Omega^{(2)}_{ijk} \om^{(2)}_{ijk}$, and
finally the length of the image as $\sum \Omega^{(3)}_{ijk}\om^{(1)}_{ijkl}$.}
\label{fig4a}
\end{figure}

The Steiner expansion
for the hyper-volume $\Omega$ of a smooth manifold $M_{4}$  reads
\be
\Omega(M^{\rho}_{4}) = \Omega(M_{4}) +\rho \cdot V(M_{4})
+ \rho^{2} \cdot S(M_{4}) + \rho^{3} \cdot A(M_{4}) +
\rho^{4}\cdot \chi(M_{4})
\label{inva4}
\ee
and by comparison with the same expression for a piecewise linear manifold
$M_{4}$ we get the discrete versions of the above invariants
\begin{eqnarray}
\Omega(M_{4}) &=& \sum_{\la ijklm\ra } v_{ijklm} \cdot 1,\label{hype}\\
V(M_{4})&=& \sum_{\la ijkl\ra } v_{ijkl} \cdot ( \pi - \alpha_{ijkl}),
\label{volu}\\
S(M_{4})&=& \sum_{\la ijk\ra } \sigma_{ijk} \cdot ( 2\pi - \sum \beta_{ijk}),
\label{areah}\\
A(M_{4})&=& \sum_{\la ij\ra } \lambda_{ij} \cdot
\omega^{(3)}_{ij}(\alpha,\beta),
\label{lineh}\\
\chi(M_{4})&=& \sum_{\la i\ra }1\cdot \omega^{(4)}_{i}(\beta), \label{disch}
\end{eqnarray}
where as before we introduce internal angles $\beta_{i}, \beta_{ij},
\beta_{ijk}$ between one, two and three simplexes and $\alpha_{ijkl}$
for the external angle between two four-dimensional simplexes,
as well as the notation $v_{ijklm}$ for the four-volume of the
four-simplex $\la ijklm\ra$ and the notation $\om^{(4)}_i$ for the
four-volume in $S^4$ of the spherical image of the vertex $\la i \ra$
(see Fig.\ \ref{fig4a}).
Note that $\omega^4_i$ is independent of the external angle $\a$.
This is in agreement with the general pattern already
mentioned which states that the coefficients to even powers
of $\rho$ are intrinsic integral invariants, while the
coefficients to the odd powers of $\rho$ contain reference to the
extrinsic geometry.
The whole four-volume (\ref{disch}) obtained by summation over
all vertices of the piecewise linear manifold $M_4$
is proportional to The Euler-Poincare
character in the same way as in the two-dimensional case.

\begin{figure}[t]
\centerline{\input{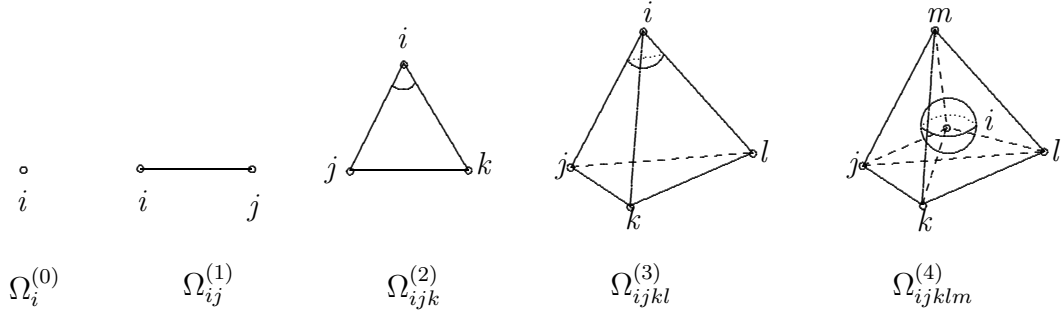}}
\caption[fig5]{Solid angles of the faces of a vertex and the
geometric measures.}
\label{fig5}
\end{figure}

To analyze  the possible new integral invariants
in four and higher dimensions, we
need to introduce a universal notation for {\it solid angles} on
the faces of a simplex and on the corresponding spherical images. We shall
use $\Omega^{(k)}$ for solid angle at a vertex of a
$k$-dimensional simplex  (see Fig.\ \ref{fig5}) and
$\omega^{(k)}$ for the solid angles on spherical images (see Fig.\ \ref{fig4a}).
All these solid angles are functions of the previously introduced angles
denoted bye $\alpha$'s and  $\beta$'s
with various indices, see Fig.\ 5.

The possible geometric measures on the spherical image of
the vertex $\la i\ra $ are: hyper-volume, volume, area and length
\cite{schneider,herglotz,allendoerfer,santalo}, (see Fig.\ 4):
\be
\omega^{(4)}_{i},~~~~~~~\sum\omega^{(3)}_{ij},~~~~~~~\sum
\Omega^{(2)}_{ijk}~\omega^{(2)}_{ijk},~~~~~~~\sum
\Omega^{(3)}_{ijkl}~\omega^{(1)}_{ijkl}.
\ee
Using these measures there is one new invariant of the ``area''
type (like  $S(M_4)$), namely
\be
\mu_{S}(M_{4})=\sum_{\la ijk\ra } \sigma_{ijk}
\cdot (~ \sum ~\omega^{(1)}_{ijkl}~) \label{fams}.
\ee
There are two new invariants  of ``length'' type (i.e. like $A(M_4)$):
\be
\mu_{A}(M_{4})=\sum_{\la ij\ra } \lambda_{ij} \cdot \left\{ \begin{array}{c}
\sum ~\omega^{(2)}_{ijk}\\
\sum~\Omega^{(3)}_{ijkl}~\omega^{(1)}_{ijkl}, \label{fama}
\end{array} \right\}
\ee
and finally three new invariants which are dimensionless,
(like $\chi(M_4)$):
\be
\mu_{\chi}(M_{4}) = \sum_{\la i\ra } 1 \cdot \left\{ \begin{array}{c}
\sum ~\omega^{(3)}_{ij}\\
\sum~\Omega^{(2)}_{ijk}~\omega^{(2)}_{ijk}\\
\sum~\Omega^{(3)}_{ijkl}~\omega^{(1)}_{ijkl} \label{famx}
\end{array} \right\}
\ee
In the next section we shall perform the dual transformation
of the above invariants and address the question of their internal
and external properties.

\subsection{Dual representations}

To get the dual form of the above invariants we combine terms
in the parentheses belonging to the same tetrahedron in the first,
third and sixth of the invariants in \rf{fams}--\rf{famx}:
\be
\left\{ \begin{array}{c}
\sigma_{ijkl} \\
\lambda_{ijkl}\\
\Omega_{ijkl}
\end{array} \right\}  = \sum_{over~tetrahedron~\la ijkl\ra }
\left\{ \begin{array}{c}
\sigma_{ijk}\\
\lambda_{ij}~\Omega^{(3)}_{ijkl}\\
\Omega^{(3)}_{ijkl}
\end{array} \right\}.
\ee
Hence, we have the following sequence of invariants constructed from
$\omega^{(1)}_{ijkl}$ associated with tetrahedron $\la ijkl\ra$:
\be
\mu_{\omega}(M_{4}) = \sum_{\la ijkl\ra } \left\{ \begin{array}{c}
v_{ijkl}\\
\sigma_{ijkl}\\
\lambda_{ijkl}\\
\Omega_{ijkl}
\end{array} \right\}  \cdot \omega^{(1)}_{ijkl}, \label{ome1}
\ee
where $\omega^{(1)}_{ijkl}=\pi - \alpha_{ijkl}$ is the angle between
two neighboring  four-dimensional simplexes having a
common tetrahedron $\la ijkl\ra $
of volume $v_{ijkl}$, area $\sigma_{ijkl}$, length $\lambda_{ijkl}$
and total internal solid angle $\Omega_{ijkl}$.

Combining the second and the fifth term in the parentheses in \rf{fams}
and \rf{famx} belonging to the same triangle
\be
\left\{ \begin{array}{c}
\lambda_{ijk}\\
\pi
\end{array} \right\}  = \sum_{over~triangle~\la ijk\ra }
\left\{ \begin{array}{c}
\lambda_{ij}\\
\Omega^{(2)}_{ijk}
\end{array} \right\}
\ee
we get integral invariants constructed from  $\omega^{(2)}_{ijk}$'s
associated to triangles $ijk$:
\be
\mu_{\omega}(M_{4}) =~~\sum_{\la ijk\ra }~~ \left\{ \begin{array}{c}
\sigma_{ijk}\\
\lambda_{ijk}\\
\pi
\end{array} \right\}~~ \cdot \omega^{(2)}_{ijk}, \label{ome2}
\ee
where $\omega^{(2)}_{ijk}=2\pi - \sum \beta_{ijk}$ is the deficit angle
on the triangle $\la ijk\ra $ which have the area $\sigma_{ijk}$, the
perimeter $\lambda_{ijk}$ and as usually the sum of internal angles
equal to $\pi$. The first invariant in this family
coincides with discrete version of the Hilbert-Einstein action
\cite{regge}, the second coincides with the linear action suggested in
\cite{savvidy2} and the last one is equal to the total deficit
angle of the whole manifold \cite{aj}.

The fourth invariant is simply equal to the volume of the spherical image
of the edge and associated to the linear term (\ref{lineh}) we have
two integral invariants constructed from the three-volume of the
spherical image $\omega^{(3)}_{ij}$ of the edge $\la ij \ra$:
\be
\mu_{\omega}(M_{4}) =~~\sum_{\la ij\ra }~~ \left\{ \begin{array}{c}
\lambda_{ij}\\
1
\end{array}
\right\}~~ \cdot \omega^{(3)}_{ij}. \label{ome3}
\ee

To understand why some of the integral invariants introduced
above are intrinsic and are independent of the embedding one
should use the Gauss-Bonnet theorem.
The Gauss-Bonnet theorem provides us with a
general relations between the volume of
the spherical image of the
vertex and the solid angles associated to the vertex.
Indeed, the Gauss-Bonnet theorem for the two-dimensional
triangulated surface can be
formulated in the form \cite{allendoerfer,santalo,santalo1,schneider}
\be
\frac{1}{4\pi}\omega^{(2)}_{i} + \frac{1}{4\pi}\Omega^{(2)}_{ijk}=
\frac{1}{2}.
\ee
Hence, summing over all vertices in the triangulation we get
\be
\chi(M_{2}) \equiv \frac{1}{2\pi}\sum_{\la i\ra}\omega^{(2)}_{i}
= \sum_{\la i\ra}\{1-\frac{1}{2\pi}\sum \Omega^{(2)}_{ijk}\}
= N_{0} - N_{2}/2 ,
\ee
where $N_{0}$ is the number of vertices and $N_{2}$ the number of
triangles on the surface $M_{2}$.

This discussion can be generalized to four dimensions:
in every vertex of the simplicial
manifold $M_{4}$ we have \cite{aw,allendoerfer,santalo,santalo1,schneider,che}
\be
\frac{3}{8\pi^{2}}\omega^{(4)}_{i} + \frac{1}{8\pi^{2}}
\Omega^{(2)}_{ijk}~\omega^{(2)}_{ijk} + \frac{1}{4\pi^{2}}
\Omega^{(4)}_{ijklm}=\frac{1}{2},
\ee
which shows that the hyper-volume $\omega^{(4)}_{i}$ on $S^4$ can be expressed
though the intrinsic  quantities.
The Euler-Poincare character is equal to
\be
\chi(M_{4}) \equiv \frac{3}{4\pi^{2}}\sum_{\la i\ra}\omega^{(4)}_{i}
= \sum_{\la i\ra} \{ 1 - \frac{1}{2\pi^{2}}\sum \Omega^{(4)}_{ijklm} \}
- \frac{1}{2}\cdot \frac{1}{2\pi}\sum_{\la ijk\ra}\omega^{(2)}_{ijk} =
N_{0}-N_{2}/2 ~ +N_{4}
\ee
and we obtain the  relation between the total deficit angle
\be
\omega^{(2)}_{tot}\equiv \frac{1}{2\pi}\sum_{\la ijk\ra}\omega^{(2)}_{ijk}
\ee
and the total solid deficit angle

\be
\Omega^{(4)}_{tot}\equiv \sum_{\la i\ra}(1-\frac{1}{2\pi^2}
\Omega^{(4)}_i) =\sum_{\la i\ra}
\{ 1 - \frac{1}{2\pi^{2}}\sum \Omega^{(4)}_{ijklm} \}
\ee
in the form  \cite{rw}
\be
\chi(M_{4}) = \Omega^{(4)}_{tot} - \frac{1}{2}\omega^{(2)}_{tot}.
\label{disvers}
\ee
The Euler-Poincare character is 
\be
\chi(M_4) = {1\over 128 \pi^2}\int_{M_4} 
dv_4 \Bigl(R^{2}_{\mu\nu\lambda\rho} -4 R^{2}_{\mu\nu} + R^{2}\Bigr),
\label{inteuler}
\eeq
and comparing with the discretized version \rf{disvers} of $\chi(M_4)$
%
it is not unnatural to associate higher powers of the
integral of curvature tensors with linear combinations of
$\Omega^{(4)}_{tot}$ and $\omega^{(2)}_{tot}$,
although we have no explicit identification of these two
terms entirely in terms of integrals of $R^2$, $R_{\m\n}^2$ etc.

From these considerations it seems that the general
discretized action in simplicial
four-dimensional gravity should be of the form mentioned in the introduction:
\bea
Action & = &\frac{1}{G}\sum_{\la ijk\ra } \sigma_{ijk} \cdot \omega^{(2)}_{ijk}
+ g_1\sum_{\la ijk\ra}\omega^{(2)}_{ijk}+    +
g_2\sum_{\la ijk\ra}\Bigl(\omega^{(2)}_{ijk}\Bigr)^2+ \nonumber\\
&&
k_1 \sum_{\la i\ra } \Bigl(1-\frac{1}{2\pi^2}\Omega^{(4)}_i\Bigr) +
k_2 \sum_{\la i\ra} \Bigl(1-\frac{1}{2\pi^2}\Omega^{(4)}_i\Bigr)^2 \label{genaction}
\eea
where the terms involving the square of the deficit angles
introduce an intrinsic rigidity into the simplicial manifolds.
This choice of discretized action with higher derivative terms
is related, but not identical to the terms suggested in \cite{hw1}.

\section{High dimensions}

It is not difficult to extend these results to high
dimensional manifolds, using the same ideas and constructions.
The extension of the formulas (\ref{hype})-(\ref{disch})
and (\ref{fams})-(\ref{famx}) or their dual form (\ref{ome1})-
(\ref{ome3}) is straightforward and have the same structure as
already encounted in four and lower dimensions.

One new aspect in higher dimensions is the appearance
of additional intrinsic integral invariants which are of 
interest in gravity and membrane theory.
The first one coincides with Hilbert-Einstein action 
\be
A(M_{5}) = \sum_{<ijkl>} v_{ijkl} \cdot \omega^{(2)}_{ijkl}
\ee
where $v_{ijkl}$ is the three-volume of the tetrahedron $<ijkl>$ 
and $\omega^{(2)}_{ijkl}$ is the deficit angle. The second 
invariant is proportional to the linear size of the manifold 
\be
A(M_{5}) = \sum_{<ij>} \lambda_{ij} \cdot \omega^{(4)}_{ij}
\ee
where $\lambda_{ij}$ is the length of the edge and $\omega^{(4)}_{ij}$
is the hyper-volume of the corresponding spherical image on $S^{5}$.
Both measures $\omega^{(2)}_{ijkl}$ and $\omega^{(4)}_{ij}$ are 
expressible in terms of intrinsic angles, as we have seen in the 
previous section, and both measures produce 
an  sequence of integral invariants in various dimensions.
The role of the deficit angle $\omega^{(2)}_{ijkl}$ is well understood 
because this measure leads to  the Euler  character of the  two 
dimensional surfaces (\ref{disc}) and  makes it possible  
to formulate discrete versions of classical  gravity 
\be
Action^{(d)} = \sum v^{(d-2)} \omega^{(2)}
\ee
where $v^{(d-2)}$ denotes the volume of the $d-2$ dimensional
sub-simplex and $\omega^{(2)}$ the deficit angle. The role 
of $\omega^{(4)}$ is very similar for four and higher dimensional manifolds
where it leads to  the Euler-Poincare character (\ref{disch})
and allows us to generate a  new sequence of invariants: 
\be
Action^{(d)} = \sum v^{(d-4)} \omega^{(4)}
\ee
which may play a role in the quantum theory of  extended objects.
It might be difficult to investigate the effect of these actions
by analytical means, but they seem well suited for use in numerical simulations.

\section{Acknowledgement}

G.K.S thanks Rolf Schneider for
helpful discussions and the Niels Bohr Institute for hospitality.
J.A. acknowledges
the support of the Professor Visitante Iberdrola Grant and
the hospitality at the University of Barcelona, where part of
this work were done.


\begin{thebibliography}{99}


\bibitem{regge}T.Regge, Nuovo Cimento 19 (1961) 558
\bibitem{ninomiya}A. Jevicki and M. Ninomiya, Phys.Rev.D33 (1986) 1634. 
\bibitem{hamber}H.W. Hamber, Nucl.Phys. B (proc.supll.) 25A (1992) 150.

\bibitem{adf} J. Ambj\o rn, B. Durhuus, J. Fr\"{o}hlich and P. Orland:\\
 Nucl.Phys. { B257} (1985) 433;  { B270} (1986) 457;
 { B275} (1986) 161-184.


\bibitem{david} F. David, Nucl.Phys. B257 (1985) 543.

\bibitem{kkm}V.A. Kazakov, I. Kostov and A.A. Migdal, Phys.Lett. B157
(1985) 295; Nucl.Phys. B275 (1986) 641.



\bibitem{kpz}V. Knizhnik, A. Polyakov and A. Zamolodchikov, Mod.Phys.Lett
             A3 (1988) 819.

\bibitem{ddk} F. David, Mod.Phys.Lett. A3 (1988) 1651; J. Distler
              and H. Kawai, Nucl.Phys. B321 (1989) 509.

\bibitem{av} 
J. Ambj\o rn, B. Durhuus and T. Jonsson. Mod.Phys.Lett.
A6 (1991) 1133.\\
J. Ambj\o rn and S. Varsted,
Phys.Lett. { B266} (1991) 285; 
 Nucl.Phys. { B373} (1992) 557.\\
M.E. Agishtein and A.A. Migdal, Mod. Phys. Lett. A6 (1991) 1863.\\
B. Boulatov and A. Krzywicki, Mod.Phys.Lett A6 (1991) 3005.\\
J. Ambj\o rn, D.V. Boulatov, A. Krzywicki and S. Varsted,
phys.Lett. { B276} (1992) 432.
N.Sakura, Mod.Phys.Lett. A6 (1991) 2613.\\
N.Godfrey and M.Gross. Phys.Rev. D43 (1991) R1749.\\
S.Catterall,J.Kogut and R.Renken. Phys.Lett. B342 (1995) 53.

\bibitem{aj}J. Ambj\o rn and J.  Jurkiewicz,
 Phys.Lett. { B278} (1992) 50. \\
M.E. Agishtein and A.A. Migdal, Mod. Phys. Lett. A7 (1992) 1039.\\
M.E. Agishtein and A.A. Migdal, Nucl.Phys. B385 (1992) 395.\\
J.Ambj\o rn,J.Jurkiewicz and C.F.Kristjansen. Nucl.Phys.
B393 (1993) 601.

\bibitem{weingarten}D.Weingarten, Nucl.Phys. B210 (1982) 229

\bibitem{many}J.Ambj\o rn,J.Jurkiewicz and C.F.Kristjansen. Nucl.Phys.
B393 (1993) 601.\\
B.Bruegmann and E.Marinari. Phys.Rev.Lett. 70 (1993) 1908.\\
\bibitem{bm}B. Bruegmann, Phys. Rev. D47 (1993) 3330.\\
S.Catterall,J.Kogut and R.Renken. Phys.Lett. B328 (1994) 277.\\
B.V. DeBakker and J. Smit, Nucl.Phys. B439 (1995) 239.\\
J. Ambj\o rn and J. Jurkiewicz , Nucl.Phys. B451 643.\\
P. Bialas, Z. Burda, A. Krzywicki and  B. Petersson, 
e-Print Archive: hep-lat/9601024 .

\bibitem{savvidy2}G.K. Savvidy and K.G. Savvidy, Interaction hierarchy.
String and quantum gravity, Mod.Phys.Lett.A (1996), Hep-th 9506184


\bibitem{steiner}J.Steiner, \"Uber parallele Fl\"achen, Gesammelte
Werke Band 2. (Berlin, 1882) S. 171-176

\bibitem{weyl}H.Weyl, Am.J.Math. 61 (1939) 461

\bibitem{ambarzum}R.V. Ambartzumian, G.K. Savvidy, K.G. Savvidy
and G.S. Sukiasian, Phys. Lett. B275 (1992) 99

\bibitem{savvidy}G.K. Savvidy and  K.G. Savvidy,
Mod.Phys.Lett. A8 (1993) 2963;\\
Int. J. Mod. Phys. A8 (1993) 3993

\bibitem{durhuus}B.Durhuus and T.Jonsson.
Phys.Lett. B297 (1992) 271


\bibitem{aw}C.B. Alendoerfer and A. Weil, Trans.Am.Soc. 55 (1943) 101-129.
\bibitem{herglotz}G.Herglotz, Abh. Math. Sem. Hansischen Univ. 15 (1943) 165

\bibitem{allendoerfer}C.B.Allendoerfer, Bull.Amer.Math.Soc. 54 (1948) 128

\bibitem{santalo}L.A.Santal\'o, Proc.Amer.Math.Soc. 1 (1950) 325

\bibitem{santalo1}L.A.Santal\'o, Revista Un.mat. Argentina 20 (1962) 79

\bibitem{savvidy1}G.K. Savvidy and K.G. Savvidy, Phys.Lett.
 B337 (1994) 333

\bibitem{schneider}R. Schneider, Geom.Dedicata 9 (1980) 111

\bibitem{che}J. Cheeger, W. M\"{u}lleer and R. Schrader, 
Comm.Math.Phys. 92 (1984) 405.

\bibitem{rw}M. Rocek and R.M. Williams, Phys.Lett. B273 (1991) 95.


\bibitem{hw1}H.W. Hamber and R.M. Williams, Nucl.Phys. B248 (1984) 392. 

\end{thebibliography}
\end{document}